\documentclass[iop]{emulateapj}
\usepackage{natbib}
\usepackage{textcomp}
\usepackage{etoolbox}
\usepackage{hyperref}
\usepackage{graphicx,color}
\usepackage{times}
\usepackage{amssymb}
\usepackage{color}
\usepackage{url}
\usepackage{amsmath}
\usepackage[english]{babel}

\begin{document}

\title{Coronal properties of the Seyfert 1 galaxy 3C 120 with {\it NuSTAR}}

\author{Priyanka Rani$^{1}$, C. S. Stalin$^{1}$}
 
\affil{$^{1}$Indian Institute of Astrophysics, Koramangala, Bangalore 560034, India. e-mail: {\color{blue}{priyanka@iiap.res.in}}\\}

\begin{abstract}
We present measurement of the cut-off energy, a proxy for
the temperature of the corona in the nuclear 
continuum of the Seyfert 1 galaxy 3C 120 using $\sim$120 ks
of observation from {\it NuSTAR}. The quality broad band spectrum
from 3$-$79 keV has enabled us to measure
the Compton reflection component (R) and to constrain the  temperature 
of the coronal plasma.
Fitting one of the advanced Comptonization models, {\it compPS} to the observed broad band spectrum
we derived the kinetic temperature of the electrons in the 
corona to be $kT_e = 25 \pm 2$ keV with Compton {\it y} parameter 
of $y = 2.2 \pm 0.1$ for a slab geometry and  
$kT_e = 26_{-0}^{+2}$ keV with a $y$ of $2.99_{-0.18}^{+2.99}$ 
assuming a spherical geometry. 
We noticed excess emission from $\sim$10$-$35 keV arising due to Compton 
reflection and a broad Fe $K\alpha$ line at 6.43 keV with an 
equivalent width of 60 $\pm$ 5 eV. The variations in count rates in the soft 
(3$-$10 keV) band is found to be more compared to the hard 
(10$-$79 keV) band  with mean fractional
variability amplitudes of 0.065$\pm$0.002 and 0.052$\pm$0.003 for the soft
and hard bands respectively. 3C 120 is known to have a strong jet, however, 
our results indicate that it is either dormant or its contribution if any 
to the X-ray emission
is negligible during the epoch of {\it NuSTAR} observation.
\end{abstract}


\section{Introduction}

Active galactic nuclei (AGN) are energetic extragalactic sources believed
to be powered by a complex physical process, namely, the  accretion of matter
on to supermassive black holes (SMBHs) residing at the centres of galaxies
\citep{1984ARA&A..22..471R}. The  matter accreted by the SMBH  forms an
optically thick, geometrically thin accretion disk 
\citep{1973A&A....24..337S}, with a temperature of about 
10$^4$$-$10$^5$ K for a SMBH of mass 10$^6$ to 10$^9$ M$_\odot$, and that emits 
predominantly in the optical/UV region of the electromagnetic spectrum.
The components of an AGN therefore  include an accretion disk surrounding the
central SMBH, an X-ray emitting corona, an obscuring torus
surrounding the accretion disk and a relativistic jet in about
15\% of AGN \citep{1995PASP..107..803U}.  While the accretion disk emits in the 
optical/UV bands, the observed hard X-ray continuum is believed to 
be caused by thermal Comptonization of
the accretion disk photons in a hot ($\sim$10$^9$ K) optically thin
corona \citep{1991ApJ...380L..51H,1994ApJ...432L..95H,1997ApJ...476..620H} 
in the radio-quiet category (those without radio jets) of 
AGN. This Comptonization process gives rise to the observed 
power-law X-ray  spectrum in AGN and the shape of this observed spectrum 
depends on the seed photon field, the kinetic temperature of the plasma 
$kT_e$, the optical depth $\tau$  and the geometry of the corona. Arguments 
in literature indicate the corona to lie between 3 $-$ 10 R$_g$ above the 
central SMBH \citep{2009Natur.459..540F}. Here, $R_g = GM/c^2$ is the 
gravitational radius
of a SBMH with mass M$_{BH}$.  In addition to
the primary power law continuum, the X-ray spectrum also contains 
a Fe K$\alpha$ line at 6.4 keV. The origin of this line, unambiguously 
detected for the first time in the source MCG-6-30-15 
\citep{1995Natur.375..659T} is attributed
to the reflection of the power law photons from the hot corona by the
relatively cold accretion disk \citep{1991MNRAS.249..352G}. This line
is now observed in many AGN \citep{2007ARA&A..45..441M}.
Therefore, X-ray observations of AGN provide very important clues to the
physical processes that happen close to the SMBH. \\

A feature in the hard X-ray spectra of AGN  that arises from the 
thermal Comptonization
process is the presence of a high energy cut off ($E_{cut}$) which is related
to the plasma electron temperature ($kT_e$) of the corona that is found to 
range between 50 to 
100 keV. 
Measurements of $E_{cut}$ 
for about hundred AGN \citep{2001ApJ...556..716P,2016AN....337..490M,2007A&A...461.1209D} are 
available as of today based on 
observations using older observatories such as the CGRO
\citep{2000ApJ...542..703Z}, {\it BeppoSAX} \citep{2002A&A...389..802P}
and the currently operating satellite 
INTEGRAL \citep{2014ApJ...782L..25M}. 
 For example, \cite{2007A&A...461.1209D}  using {\it BeppoSAX} data in the energy range 2 $-$ 100 keV 
has provided $E_{cut}$ measurements for 34 sources and lower limits
for another 54 sources. 
Also, \cite{2014ApJ...782L..25M} using data from XMM, INTEGRAL and
{\it Swift} has provided $E_{cut}$ values for 26 AGN and lower limits for 11 others.
Most of these measurements have large error bars, 
which may be due to  
the quality of the data itself 
as well as the degeneracy that is known to prevail between $E_{cut}$ 
and other physical properties of the sources such as the slope of
the primary X-ray power law, and the amount of radiation that are
Compton up scattered by the circumnuclear material \citep{2016AN....337..490M}.

One of the important physical prameters of the Comptonizing corona in an AGN is its plasma temperature $kT_e$. The value of 
$kT_e$ in many instances is indirectly obtained by fitting simple phenomnological models such as the 
cut-off power law to the observed X-ray spectrum to find $E_{cut}$ and subsequently $kT_e$ is deduced
using certain approximations. However, $kT_e$ values are available for many
AGN based on physical model fits to AGN spectra \citep{2001ApJ...556..716P,2016MNRAS.458.2454L}.
 Disentangling the different spectral components that are present in the 
X-ray spectra of AGN and constraining their characteristic parameters require
high S/N data covering a wide energy range.
These limitations are now overcome to a large extent owing to the availability
of the focussing hard X-ray telescope, 
the Nuclear Spectroscopic Telescope Array, {\it NuSTAR} 
\citep{2013ApJ...770..103H}, due to its 
high sensitivity and  wide energy coverage from 3 $-$ 79 keV.
 However, {\it NuSTAR} too can provide precise measurements only for sources
with $E_{cut}$ $<$ 150 keV \citep{2014ApJ...782L..25M}. Measurements beyond this from {\it NuSTAR}
will not be reliable due to its own lack of effective area beyond 79 keV.
Using {\it NuSTAR}, $E_{cut}$ as well as the 
more physical coronal plasma temperature $kT_e$ has been measured in about a 
dozen AGN \citep{2014ApJ...781...83B,2014MNRAS.440.2347M,
2015ApJ...800...62B,2014ApJ...794...62B,2015MNRAS.447.3029M,2015ApJ...814...24L,
2017MNRAS.466.4193T,2017MNRAS.465.1665K, 2017MNRAS.470.3239P,2017ApJ...841...80L,2017MNRAS.468.3489K}
and also summarized by \cite{2016AN....337..490M}. 
Though {\it NuSTAR} observations have enabled us to characterise the 
coronal properties of about a dozen AGN, to have a complete knowledge of 
the geometrical and physical properties of the X-ray corona in AGN,
 there is a need to 
extend such studies on the X-ray properties to a large sample of AGN.

In this paper, we present the results  of our analysis of X-ray data 
on 3C 120 observed by {\it NuSTAR} for a total duration of about 120 ks.
3C 120 is a X-ray bright Seyfert 1 galaxy at $z$ = 0.033
\citep{1967ApJ...149L..51B} and  
having a black hole mass of $5.6 \times 10^{7}$ M$_{\odot}$ \citep{2015PASP..127...67B}. 
It is also classified as a broad line radio galaxy (BLRG) by 
\cite{1987ApJ...316..546W}. It has a radio morphology similar to 
the FRI category of AGN \citep{1974MNRAS.167P..31F}.
Its one sided jet has an inclination to the line of sight 
of $\sim$ 14$^{\circ}$ \citep{1998ApJ...505..577E} which is  based on 
the  apparent superluminal speed $\beta_{app}$ reported by
\cite{1989LNP...334....3Z}. The jet is known to extend on scales up to
100 kpc \citep{1987ApJ...316..546W}. It has been found to be variable in
X-rays. A broad Fe K$\alpha$ line well fitted
by a Gaussian with a $\sigma$ of 0.8 keV and having an equivalent width of
400 eV has been found from ASCA observations \citep{1985ApJ...290..130H}.
It has not been detected in $\gamma$-rays by the Compton Gamma Ray 
Observatory (CGRO,\citealt{1993ApJ...416L..53L}). However, using data from 
the Oriented Scintillation Spectroscopy Experiment (OSSE,\citealt{1993ApJS...86..693J}) 
on board CGRO, \cite{1998MNRAS.299..449W} found the
presence of a spectral break in 3C 120 between X-rays and soft $\gamma$-rays.
3C 120 was detected in {\it Fermi}
using the first 15 months of data \citep{2010ApJ...720..912A}, but
not detected in the second {\it Fermi}-LAT catalog (2FGL, 
\citealt{2012ApJS..199...31N}) and the third {\it Fermi}-LAT
catalog (3FGL, \citealt{2015ApJS..218...23A})
indicating that the source is variable in the high energy $\gamma$-rays.
Using 180 and 365 days binning on the data obtained between
August 2008 - December 2013, \cite{2015A&A...574A..88S} found 
$\gamma$-ray flux variations. Using 5 days binned light curve
\cite{2015ApJ...799L..18T} noticed that 3C 120 was detected only 
at certain epochs.

In this work, we focus on both the timing and broad band (3$-$79 keV) spectral 
analysis. Though 3C 120 has
been studied for flux variability as part of the analysis of a large sample
of AGN by \cite{2017MNRAS.466.3309R}, analysis of the broad band
X-ray data to constrain its coronal properties using {\it NuSTAR} has 
been carried out for the first time. However, from {\it BeppoSAX}
observations, \cite{2001ApJ...551..186Z} have estimated a 
$E_{cut}$ of 100$-$300 keV. Also, \cite{1998MNRAS.299..449W} using the average OSSE spectrum 
together with ASCA data reported a $E_{cut}$ of 130$_{-40}^{+150}$ keV. Using data 
from several telescopes \cite{2016MNRAS.458.2454L} obtained a value of $kT_{e} = 176_{-23}^{+24}$ keV. 
This paper is organised as follows.
 In Section 2 we report on the {\it NuSTAR} observations and data reduction. 
In Section 3 we present the analysis of the 
{\it NuSTAR} data. The results are discussed in Section 4 followed by
the summary in the final section.


\section{Observation and Data reduction}
3C 120 was observed by {\it NuSTAR} \citep{2013ApJ...770..103H} on 
6$^{th}$ 
February 2013 (ObsID 60001042003) for 127 ks in the 3$-$79 keV band. 
The data was reduced using the {\it NuSTAR} Data Analysis Software 
package {\it NuSTARDAS} v.1.6.0 distributed by the High Energy Astrophysics 
Archive Research Center (HEASARC). We generated the cleaned and screened 
event files taking also into account the passage of the 
satellite through the South Atlantic Anomaly using the {\it nupipeline} task and using CALDB 20161207. A 
circular region of radius $60''$ was taken centered
 on the peak of the source image to extract 
the source spectrum and light curve. To extract the background spectrum
and light curve we again selected a circular region of
 radius  $60''$ away from the 
source  on the same detector. Light curves were generated with 300 seconds 
binning in the 3$-$79 keV band for both the focal
plane modules, FPMA and FPMB. They were further 
divided into soft (3$-$10 keV) and hard (10$-$79 keV) bands. To get the final 
light curves, the count rates from the two modules FPMA and FPMB were 
combined using the {\it lcmath} task in FTOOLS V6.19. 

We generated the source and background spectra and response files in the 
energy range of 3$-$79 keV in each focal plane module FPMA and FPMB using the 
{\it nuproducts} package available in {\it NuSTARDAS}. Instead of combining 
the spectra or responses from FPMA and FPMB, we fitted them simultaneously, 
allowing the cross normalization for both modules to vary freely in all fits. 
We used {\it XSPEC} (version 12.9.0; \citealt{1996ASPC..101...17A}) for 
all spectral analysis. During the various model fits in {\it XSPEC}, we 
used \cite{1989GeCoA..53..197A} set of solar abundances and the \cite{1992ApJ...400..699B} photoelectric cross 
sections. The $\chi^2$ minimization technique in {\it XSPEC} was used 
to get the best model description of the data and all errors were 
calculated using $\chi^2$=2.71 criterion, i.e., 90\% confidence level.


 To check for any possible effects
on our choice of the background region used to get the light curves and the
source spectrum, we also
used a background found in an annular region around the source having an
inner radius of 50 pixels and an outer radius of 80 pixels (1 pixel = 2.46 arcsec). The mean
count rates of the source light curve in the 3$-$79 keV band using the background taken from the circular 
region away from the source and using the annular region surrounding the source
are 3.88 $\pm$ 0.20 and 3.84 $\pm$ 0.20 respectively. The choice of the background thus has neglible
effect on the light curves presented here. Similarly, the average fluxes
of the source in the 3$-$79 keV band obtained from a simple power law fit to the spectrum  
are (1.372 $\pm$ 0.007) $\times$ 10$^{-10}$ and  (1.237 $\pm$ 0.007) $\times$ 10$^{-10}$ erg s$^{-1}$ cm$^{-2}$ respectively
using the background taken from the circular region away from the source and the annular region surrounding the source. 
The difference in the fluxes deduced from the spectra using two choices of the background
selection is less than 10\%. 
Thus, the selection of background region has no significant impact on the spectral and timing analysis carried out on 
3C 120.

\section{Analysis of the data}
\subsection{Flux variability}
\begin{figure}
\centering
\hspace*{-0.8cm}\includegraphics[width=11cm,height=12cm]{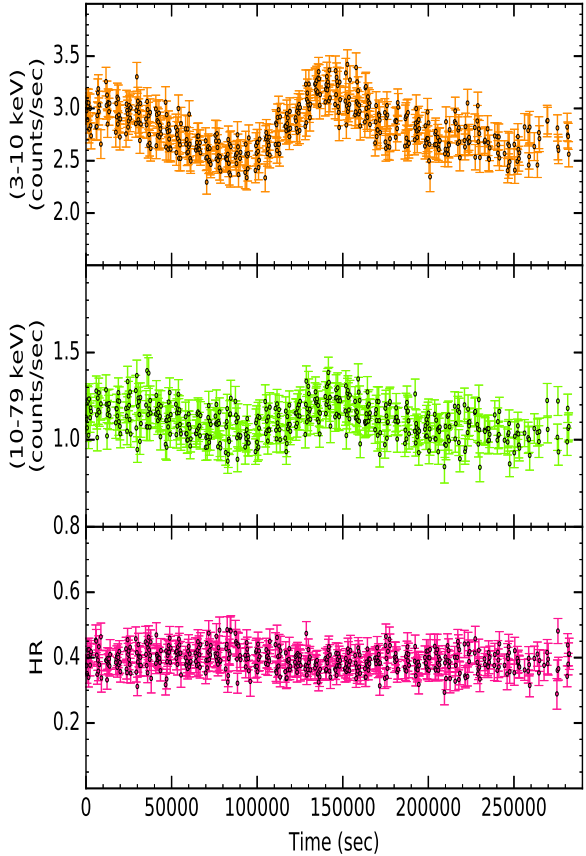}
\caption{\label{spectra1}Light curves in counts/sec for the  soft (top) and 
hard (middle) bands. The zero point of the time axis is 2013-02-06T23:51:07}. 
The variation of HR is shown on the bottom panel.
\end{figure}

\begin{figure}
\centering
\hspace*{-0.8cm}\includegraphics[height=8cm,width=12cm]{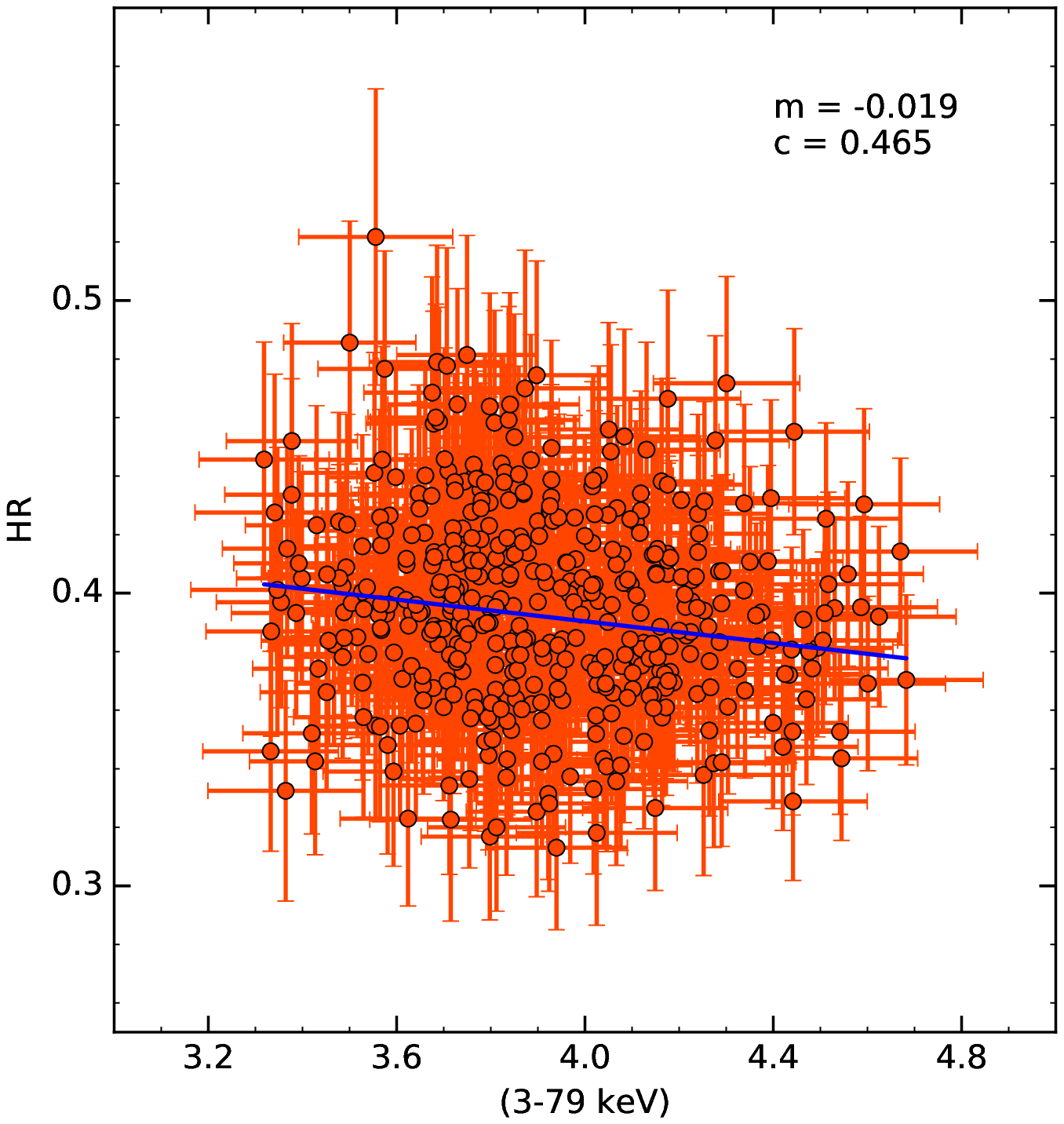}
\caption{\label{spectra2} Correlation between HR and count rate variation}
 in the 3$-$79 keV band. The solid blue line is 
the linear least squares fit to the data.
\end{figure}

In Figure \ref{spectra1} we show the light curves in two energy bands 3$-$10 keV and
10$-$79 keV as well the hardness ratio (HR) curve. The HR is evaluated as
the ratio between the count rates in the  10$-$79 keV to 3$-$10 keV bands. The source
is found to show variations in both the soft and hard bands. To 
characterise the flux variations from observed count rates, we used the 
fractional root mean
square variability amplitude (F$_{var}$) which gives an estimate of the 
intrinsic variability amplitude relative to the mean count rate 
\citep{2002ApJ...568..610E,2003MNRAS.345.1271V}.   We found
F$_{var}$ values of 0.065 $\pm$ 0.002 and 0.052 $\pm$ 0.003 respectively for the 
soft and hard bands. To characterise the presence of spectral variations
if any, we calculated the HR and show in Figure \ref{spectra2} the plot of HR against
flux variations (in the units of counts/sec) in the total (3$-$79 keV) band. Fit of a linear function 
of the form HR = m $\times$ counts/sec (3$-$79) keV + c, that also
takes into consideration the errors in both HR and the 3$-$79 keV count rates, 
gave values of $-$0.019 $\pm$ 0.005 and 0.465 
$\pm$ 0.022 for m and c respectively. The spectrum is thus found not to change
with the brightness of the source.

\subsection{Spectral analysis}
Analysis of the spectrum of 3C 120 along with  model fittings was
carried out using the {\it XSPEC} package.
To find the best fit models, the $\chi^{2}$ minimization technique 
in {\it XSPEC} was used.

Analysis of X-ray spectra are generally attempted by fitting simple 
phenomenlogical models such as the simple power law model as well as
the power law with exponential cut-off. Using power law fits, \cite{retco} 
obtained a value of $\Gamma$ = 1.85 $\pm$ 0.01. Similarly, using the 
{\it pexrav} model, the same authors obtained value of the high energy
cut-off ($E_{cut}$) to be 83$^{+10}_{-8}$ keV. 
However, these models are too simplistic. 
Also, the power law with an exponential cut off 
that is incorporated in 
{\it pexrav} has limitations in its approximation of the steep decline of the 
true high energy cut off 
\citep{2003MNRAS.342..355Z,2015MNRAS.451.4375F,2016MNRAS.458.2454L}. 
Thus, both the models are too simplistic to apply to the good quality of the 
data from NuSTAR that is analysed here. Therefore, the following more physical models
were fit to the data.

\subsubsection{CompTT model}
\begin{figure}
\centering
      \includegraphics[width=8cm,height=8cm]{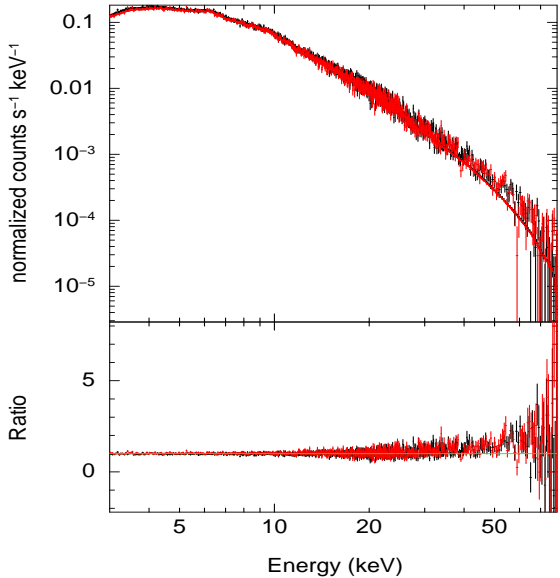}
\caption{\label{model3sphere} The figure shows the 
observed spectrum (Normalized counts/sec versus Energy) along with
the fitted model TBabs*ZTBabs*(zgauss+compTT+refl(compTT))(for a spherical geometry) in 
FPMA(black) and FPMB(red). The ratio of observations to the fitted model is
also shown for FPMA (black) and FPMB (red).}
\end{figure}

\begin{figure}
\centering
      \includegraphics[width=8cm,height=8cm]{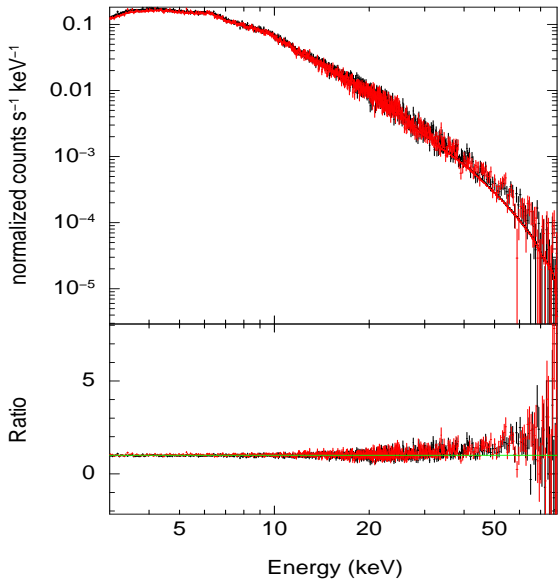}
\caption{\label{model3slab} Same as in Figure \ref{model3sphere} except for the slab geometry.}
\end{figure}

We used the Comptonization model ({\it CompTT}; 
\citealt{1994ApJ...434..570T}) convolved  with a reflection component so as to 
get the coronal parameters. 
This model has the form {\it TBabs $\times$ zTBabs $\times$ 
(zgauss$+$CompTT$+$refl(CompTT))}. 
The first component of this model {\it TBabs} \citep{2000ApJ...542..914W} 
includes galactic absorption, with
the galactic neutral hydrogen column density frozen to the value of 
$1.11 \times 10^{21}$ cm$^{-2}$ obtained from \cite{1990ARA&A..28..215D} 
using the nH tool in HEASARC\footnote{https://heasarc.gsfc.nasa.gov/cgi-bin/Tools/w3nh/w3nh.pl} and the second component 
{\it zTBabs} represents absorption intrinsic to the host galaxy of the source. 
Redshift was fixed to $z=0.033$ and the column density {\it zTBabs} was kept as a free parameter 
in the fitting. 
The {\it CompTT} component in this model assumes 
a geometry for the corona (a slab or spherical) and models the intrinsic 
coronal continuum, and {\it refl} convolves it with reflection features 
\citep{2015ApJ...800...62B}. 
For slab geometry the reduced 
$\chi^{2}$ was 0.986 ($\chi^{2}/\nu=3734/3788$) and for spherical
geometry too it was 0.986 ($\chi^{2}/\nu=3734/3788$). For the slab 
geometry, 
we found the mean value of $kT_e = 9_{-3}^{+2}$ keV and $\tau = 2.4_{-1.1}^{+0.6} $ considering 
primary and reflected emission. 
at the 90\% confidence. For the spherical geometry 
the best fit yielded the mean value of $kT_e = 16_{-7}^{+6}$ keV 
and $\tau = 5.1_{-0.4}^{+0.6}$ for 
primary and reflected emission at the 90\% confidence. The {\it CompTT} model 
gave huge error bars in the normalization constant.
 The observed spectrum along with the 
fit and residuals are shown in 
Figure \ref{model3sphere} for the spherical geometry 
and Figure \ref{model3slab} for the slab 
geometry.
The best fit parameters and their errors at 
90\% confidence levels are given in Table \ref{log1}.
The 2$-$10 keV flux determined from the fit is (5.19 $\pm$ 0.01) 
$\times$ 10$^{-11}$ erg cm$^{-2}$ sec$^{-1}$. 
This gives an unabsorbed luminosity of 
(1.29 $\pm$ 0.01) $\times$ 10$^{44}$ erg sec$^{-1}$. Using the 
bolometric correction of 20.6 $\pm$ 0.1 found by 
\cite{2009MNRAS.392.1124V} we obtained a bolometric luminosity of
(26.656 $\pm$ 0.001) $\times$ 10$^{44}$ erg sec$^{-1}$. For a BH mass 
of 5.6 $\times$ 10$^{7}$ M$_{\odot}$ \citep{2015PASP..127...67B},
 the calculated Eddington accretion rate is 
$\lambda_{Edd}$ = $L_{bol}/L_{Edd}$ = 0.353, where the Eddington luminosity,$L_{Edd}$ = 
1.36 $\times$ 10$^{38}$ (M$_{BH}$/M$_{\odot})$ erg sec$^{-1}$. This is similar to the value of $\lambda_{Edd}$ = 0.352 
found by \cite{2016MNRAS.458.2454L} using data from many telescopes including {\it INTEGRAL}.\\

\subsubsection{CompPS}
Though {\it CompTT} model well represents the observed spectrum, this 
has limitations such as its simplistic treatment of the seed photons that participate
in the Comptonization process. We therefore fit the spectrum using one of 
the most
advanced Comptonization models available in {\it XSPEC} namely 
{\it CompPS} \citep{1996ApJ...470..249P}. {\it CompPS} that produces
the continuum through thermal Comptonization processes
incorporates proper treatment of the Comptonization process through
exact numerical solution of the radiative transfer equations.  It also
offers several choices for the geometries. In the fitting process, when
all the parameters were kept free, the fitting failed to converge. Therefore, to avoid  
non-convergence owing to the presence of many free parameters in {\it CompPS} fitting, we fixed 
the centroid energy and deviation of the Fe K$\alpha$ line to be 
6.43 keV and 0.29 keV respectively obtained from {\it CompTT} for a slab
geometry. The parameters obtained from the fit along with their
associated errors including the Compton {\it y} parameter ($ y = 4 \tau \frac{KT}{m_ec^2}$, where 
$\tau$ is the Thomson optical depth \citep{2000ApJ...542..703Z} and the normalization (N$_{CompPS}$)
are given in Table \ref{log1}. 
To compare our results with {\it CompTT}, we used 
{\it CompPS} model only for slab and spherical geometries. The seed photons 
were assumed to be 10 eV.
 The observed spectra along with the 
fit and residuals are shown in 
Figure \ref{model45} for the spherical and the slab 
geometries. 
For slab geometry we obtained a reduced 
($\chi^{2}/\nu=3773/3790$) of 0.996, whereas, for the spherical geometry the reduced
($\chi^{2}/\nu=3783/3790$) was 0.998. For slab and spherical geometry, we found 
R values of 0.80$_{-0.10}^{+0.11}$ and 0.43$\pm$0.06 respectively. 
We found that 
the $kT_e$ values obtained from {\it CompPS} model is larger than that
obtained from {\it CompTT} for both spherical and slab geometries.

\subsubsection{EQPAIR}
\begin{figure*}
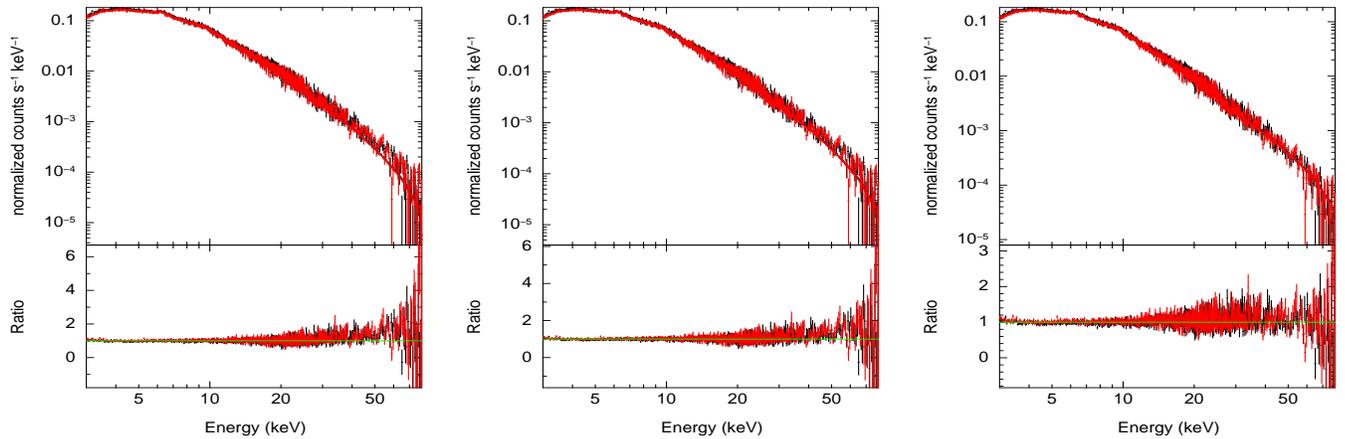

\centering
\hbox{
      \includegraphics[width=6cm,height=6cm]{Figures2/sl10ev.ps}
      \includegraphics[width=6cm,height=6cm]{Figures2/sp10ev.ps}	
      \includegraphics[width=6cm,height=6cm]{Figures2/re14.ps}
     }
\caption{\label{model45} The left panel shows the 
observed spectrum (Normalized counts/sec versus Energy) along with
the fitted model TBabs*zTbabs*(zgauss+compPS)(for a slab geometry) for the 
FPMA(black) and FPMB(red). Middle panel is same as first except for the spherical geometry.
The right panel shows the observed spectrum fitted with {\it EQPAIR} (TBabs*zTbabs*(zgauss+EQPAIR)).}
\end{figure*}

The models used above to fit the {\it NuSTAR} data of 3C 120
assumes that the electrons involved in the Comptonization process are 
thermal electrons with a Maxwellian energy distribution.
 However, hybrid models for the corona that involves the contribution
of both thermal and non-thermal electrons have been applied to
AGN. For example, in NGC 4151, the contribution of non-thermal 
electrons is found to be less than 15\% \citep{1996MNRAS.283..193Z,1997ApJ...482..173J}. 
The non-thermal fraction can also be as large as 30\% \citep{2017MNRAS.467.2566F}.
3C 120 is 
known to have a jet emission \citep{1998ApJ...505..577E} and
it is likely the observed X-ray emission is a combination of various
components.  
We therefore model the spectrum 
with the {\it EQPAIR} model \citep{1999ASPC..161..375C}, the most
advanced Comptonization model in XSPEC. This model
evaluates the emission spectrum resulting from Comptonization, Coulomb
collisions and pair production. This model  can 
treat the  Comptonization for a different nature of plasma (thermal, 
non-thermal and hybrid plasma) and even incorporates Compton hump from cold 
reflection. This is because in {\it EQPAIR}, only electrons with optical depth
($\tau_p$) are accelerated with total power characterised by the
compactness parameter $l_h$ (such that $l_h = L_h \sigma_T/Rm_ec^2$, 
where $L_h$ is the luminosity or the power supplied to the  electrons
in the Comptonization region  and R  
the size of the Comptonization region \citep{2003MNRAS.342.1041D} that is split
between thermal distribution with power ($l_{th}$) and non-thermal 
distribution with power $l_{nth}$ and $l_h = l_{nth} + l_{th}$. These electrons then cool either through
Compton scattering of soft photons or through Coulomb collisions. The parameter
that plays an important role in characterising the overall spectral shape in 
{\it EQPAIR} is the parameter $l_h$/$l_s$. The soft compactness parameter ($l_s$)
refers to the luminosity of the soft photons that is injected into the
corona and the hard compactness parameter $l_h$ refers to the power
supplied to the accelerated electrons in the source. To model the observed spectrum with {\it EQPAIR}, we fixed the centroid 
energy of the Fe K$\alpha$ emission line to the best fit 
value found from {\it CompTT} above for a slab geometry. However, in {\it EQPAIR}
model the geometry is spherical and the photons are induced 
homogeneously throughout the spherical cloud. We assumed that the 
input source of soft photons  
in {\it EQPAIR} is {\it diskpn}, a black body spectrum \citep{1999MNRAS.309..496G} with a peak temperature of 
10 eV . The seed photon distribution can be modified in the model
by changing $l_s$ which for this model fitting was fixed to 10. 
The inclination was fixed to 17 degrees and the iron abundance was taken to be
solar. 
For the purpose of this modelling we  
considered the accelerated particles to be  electrons from the thermal pool.
The best fit parameters are given in Table \ref{log1}. The observed
and fitted spectra along with the residuals are given in Figure \ref{model45}. Similar to {\it CompPS} and {\it CompTT} models, 
the fit of the data with {\it EQPAIR} model too provides a good description of the 
data with a reduced $\chi^2$ of 1.047($\chi^{2}/\nu=3972/3791$). However, the temperature is not among
the default output parameters returned by {\it EQPAIR} though it is calculated
in the model. We therefore used the {\it chatter} command ({\it chatter} level 
= 15) \citep{1999ASPC..161..375C}  and obtained $kT_e$ = 23. The  
error in $kT_e$ was obtained using $\chi^{2}$ minimization technique at the 90 \% significance level. Thus, using the
{\it EQPAIR} model we found $kT_e$ = 23$^{+1}_{-7}$ keV. 
 We found the best fit ratio of the hard to the 
soft compactness parameter, $l_h/l_s$ = $0.90 \pm 0.32$. This points to
 similar power in the irradiating soft photons that enter the source region and in the heating of the 
electrons. The {\it EQPAIR} model yields the value of ionization parameter 
of the reflector to be $\xi =$ 5.14 $\pm$ 11.15. The 
electron optical depth obtained by the fit was $\tau_p$ = $0.60 \pm 0.08$. The ratio $l_{nth}/l_h$, 
which gives the
fraction of power supplied to energetic particles that goes into accelerating 
non-thermal particles was found to be $0.78 \pm 0.10$. The value
of $l_{nth}/l_h$ is zero for a purely thermal model, while it is unity for a purely non-thermal
model. Though $l_{nth}/l_h$ obtained from the fit deviates much from zero, the detection of 
a high energy cutoff in the {\it NuSTAR} spectrum not much beyond the sensitivity of {\it NuSTAR} 
\citep{retco} and the non-detection of the source in $\gamma-$rays suggest that Comptonization 
by non-thermal electrons if any is non-significant.
 The $\chi^2$ from
EQPAIR model fit is poorer compared to CompPS and CompTT, though, the $kT_e$ 
value from EQPAIR model fit agrees to that obtained from CompPS. 
Thus, it is likely that in 3C 120, for the observations analysed here, the 
electrons involved
in the Comptonization process are predominantly thermal.



\begin{table*}
\caption{\label{log1}Best fit parameters and errors (90\% confidence) obtained from spectral 
fitting for different models. In CompPS model the parameters marked with  * were fixed to the best
fit values obtained from CompTT for a slab geometry. The errors in the parameters obtained from 
 EQPAIR are the 1 $\sigma$ error returned by the model fits.}
\centering
\begin{scriptsize}
\begin{tabular}{@{}llll@{}}
\hline
Model Name                         &   Parameter            &  Parameter                    &  ${\chi^{2}/dof}$ \\
                                     &   Name (units)            &  Values                       &                   \\
\hline
CompTT                           &                            &                   &  \\
 TBabs*zTbabs*(zgauss+CompTT+refl(CompTT))  &  $E$ (keV)        &  6.43 $\pm$ 0.06    &0.986    \\[1ex]
 (Spherical geometry)            &  $\sigma$ (keV) &  0.29$_{-0.09}^{+0.10}$    &    \\[1ex]
                            &  N$_{zgauss}$ $\times$ 10$^{-5}$ & 5.50$_{-1.01}^{+1.05}$    &    \\[1ex]
                            &  mean  $kT_e$ (keV)  &  16$_{-7}^{+6}$ &    \\[1ex]
                            &  mean  $\tau$    &  5.1$_{-0.4}^{+0.6}$    &    \\[1ex]
                            &  N$_{CompTT}$ $\times$ 10$^{6}$ & 3.58$_{-2.99}^{+234}$    &    \\[1ex]
                            &  $R$        &  0.29 $\pm$ 0.07    &    \\[1ex]
                            &  N$_{refl(CompTT)}$ $\times$ 10$^{-2}$ & 2.82$_{-2.67}^{+2.63}$    &    \\[1ex]
\hline

CompTT                          &                            &                   &  \\
 TBabs*zTbabs*(zgauss+CompTT+refl(CompTT))  &  $E$ (keV)        &  6.43 $\pm$ 0.06    &0.986    \\[1ex]
 (Slab geometry)            &  $\sigma$ (keV) &  0.29$_{-0.09}^{+0.10}$    &    \\[1ex]
                            &  N$_{zgauss}$ $\times$ 10$^{-5}$ & 5.41$_{-1.05}^{+1.63}$    &    \\[1ex]
                            &  mean $ kT_e$ (keV)  &  9$_{-3}^{+2}$    &    \\[1ex]
                            &  mean $ \tau$    &  2.4$_{-1.1}^{+0.6} $    &    \\[1ex]
                            &  N$_{CompTT}$ $\times$ 10$^{7}$ & 1.17$_{-1.16}^{+590}$    &    \\[1ex]
                            &  $R$        &  0.30$_{-0.08}^{+0.07}$    &    \\[1ex]
                            &  N$_{refl(CompTT)}$ $\times$ 10$^{-2}$ & 2.79$_{-0.23}^{+0.25}$    &    \\[1ex]
\hline

CompPS                          &                            &                   &  \\
TBabs*zTbabs*(zgauss+CompPS)       &  $E$ (keV)        & 6.43$^{*}$    &0.996    \\[1ex]
(Slab geometry)               &  $\sigma$ (keV) &  0.29$^{*}$    &    \\[1ex]
                           &  $kT_e$ (keV)  &  25 $\pm$ 2    &    \\[1ex]
                           &  Compton $y$ parameter    &  2.2 $\pm$ 0.1    &    \\[1ex]
                           &  $R$        &  0.80$_{-0.10}^{+0.11}$    &    \\[1ex]
                           &  $\xi$ & 2.36$\times$ 10$^{-3}$$_{-0.003}^{+0.237}$  &    \\[1ex]
                           &  N$_{CompPS}$ $\times$ 10$^{+8}$ & 3.23 $\pm$ 0.03    &    \\[1ex]
\hline
CompPS                          &                            &                   &  \\
TBabs*zTbabs*(zgauss+CompPS)  &  $E$ (keV)        & 6.43$^{*}$    &0.998    \\[1ex]
 (Spherical geometry)           &  $\sigma$ (keV) & 0.29$^{*}$    &    \\[1ex]
                           &  $kT_e$ (keV)  &  26$_{-0}^{+2}$    &    \\[1ex]
                           &  Compton $y$ parameter    & 2.99$_{-0.18}^{+2.99}$    &    \\[1ex]
                           &  $R$        & 0.43 $\pm$ 0.06     &    \\[1ex]
                           &  $\xi$ & 2.13$\times$ 10$^{-3}$$_{-0.002}^{+0.161}$  &    \\[1ex]
                           &  N$_{CompPS}$ $\times$ 10$^{+7}$ & 5.26$_{-0.06}^{+0.49}$    &    \\[1ex]
\hline
EQPAIR                          &                            &                   &  \\
TBabs*zTbabs*(zgauss+EQPAIR)       &  $l_{h}/l_{s}$       & 0.90 $\pm$ 0.32    &1.047    \\[1ex]
                           &  $l_{nt}/l_{h}$ & 0.78 $\pm$ 0.10    &    \\[1ex]
                                   &  $kT_e$ (keV)  & 23$^{+1}_{-7}$  &       \\[1ex]
                           &  $\tau_{p}$       & 0.60 $\pm$ 0.08    &    \\[1ex]
                           &  $R$   &  0.19 $\pm$ 0.03    &    \\[1ex]
                           &  $\xi$ & 5.14 $\pm$ 11.15  &    \\[1ex]
                           &  N$_{EQPAIR}$ $\times$ 10$^{-3}$ & 0.69 $\pm$ 0.03    &    \\[1ex]
\hline
\end{tabular}
\end{scriptsize}
\end{table*}

\begin{table*}
\caption{\label{log2}Summary of the physical parameters of the AGN
taken from literature which were obtained from {\it NuSTAR} data. Column 1 gives the name of the source, column 2 gives
the type of the source, column 3 gives the cut-off energy in keV, column 4 
is the plasma temperature of the corona for a spherical geometry, column 5 is 
the photon index, column 6 is the 
logarithm of the black hole mass in solar mass units, column 7 is 
Eddington  
accretion rate and column 8 gives the references to the 
sources from where their physical parameters were taken}
\centering
\begin{tabular}{@{}llllllll@{}}
\hline
Name          &  type &  $E_{cut}$ (keV)	&  $kT_e^{a}$ & $\Gamma$  &  log (M$_{BH}$)  & $\lambda_{Edd}$ & Reference\\[1ex]
\hline
3C 382          &  BLRG/Sy 1     & 214$_{-63}^{+147}$ & 330 $\pm$ 30         & 1.68$_{-0.02}^{+0.03}$ &  9.2 $\pm$ 0.5 &  0.109  &  1        \\[1ex]
3C 390.3        &  BLRG/Sy 1     & 116$_{-8}^{+24}$   & 16$_{-2}^{+4}$       & 1.70 $\pm$ 0.01 &  8.4 $\pm$ 0.1 &  0.240  &  2        \\[1ex]
IC 4329A        &  Sy 1.2       & 186$_{-14}^{+14}$  & 50$_{-3}^{+6}$       & 1.73 $\pm$ 0.01 &  6.8 $\pm$ 0.2 &  0.082  &  3       \\[1ex]
MCG-5-23-16     &  NELG/Sy 2    & 116$_{-5}^{+6}$    & 25 $\pm$ 2           & 1.85 $\pm$ 0.01 &  7.8 $\pm$ 0.2 &  0.031  &  4        \\[1ex]
NGC 5506        &  Sy 1.9       & 720$_{-190}^{+130}$ & ---                 & 1.91 $\pm$ 0.03 &  8.0 $\pm$ 0.2 &  0.013  &  5        \\[1ex]
J2127.4+5654    &  NLSy1        & 108$_{-10}^{+11}$  & 53$_{-26}^{+28}$     & 2.08 $\pm$ 0.01 &  7.2 $\pm$ 0.0 &  0.090  &  6        \\[1ex]
GRS 1734-292    &  Sy 1         & 53$_{-8}^{+11}$    & 12.1$_{-1.2}^{+1.8}$ & 1.65 $\pm$ 0.05 &  8.5 $\pm$ 0.1 &  0.033  &  7        \\[1ex]
4C 74.26        &  BLRG/Sy 1    & 183$_{-35}^{51}$   & 46$_{-11}^{+25}$    & 1.84$_{-0.02}^{+0.03}$   & 9.6 $\pm$ 0.5 &  0.037 &  8        \\[1ex]
Ark 564         &  NLSy1        & 42$_{-3}^{+3}$     & 15$_{-1}^{+2}$      & 2.27 $\pm$ 0.08     &  6.4 $\pm$ 0.5  &  1.10 & 9   \\[1ex]
QSO B2202-209   &  RQQ          &  153$_{-54}^{+103}$ & 42 $\pm$ 3           & 1.82 $\pm$ 0.05     &  9.1 $\pm$ 0.2  &  1.15  & 10  \\[1ex]
NGC 5273        &  Sy 1.5       & 143$_{-40}^{+96}$  & 57$_{-11}^{+18}$    & 2.27 $\pm$ 0.08    &  6.4 $\pm$ 0.5   &  1.1    & 11  \\[1ex]
3C 120          & BLRG/Sy 1     & 83$_{-8}^{+10}$    & 26$_{-0}^{+2}$     & 1.87 $\pm$ 0.02 &  7.7 $\pm$ 0.1 &  0.353  & This work \\[1ex] \hline
\end{tabular}

References: 1:\cite{2014ApJ...794...62B}, 2:\cite{2015ApJ...814...24L}, 
3:\cite{2014ApJ...788...61B}, 4:\cite{2014ApJ...788...61B}, 
4:\cite{2015ApJ...800...62B}, 5:\cite{2015MNRAS.447.3029M}, 
6:\cite{2014MNRAS.440.2347M}, 7:\cite{2017MNRAS.466.4193T},
8:\cite{2017ApJ...841...80L,2002ApJ...579..530W},
9:\cite{2007MNRAS.381.1235V, 2017MNRAS.468.3489K}
10.\cite{2017MNRAS.465.1665K}
11.\cite{2017MNRAS.470.3239P}
$^a$ the quoted $kT_e$ values for the sources except NGC 5273 pertain to spherical geometry. For 
3C 120 the quoted value is that returned by {\it compPS} model fit.
\end{table*}

\section{Discussion}

\subsection{Coronal properties}
The availability of high quality {\it NuSTAR} data from observations of 
about 120 ks has enabled the determination of the coronal properties of 
3C 120.  The time averaged spectrum covering the 
3$-$79 keV band, when fitted with the  phenomenological power law model gave 
the continuum power law index of $\Gamma$ = 1.85 $\pm$ 0.01
 \citep{retco}. However, values of 1.70 and 2.08 were found from 
XMM \citep{2009MNRAS.392.1124V} and INTEGRAL \citep{2016MNRAS.458.2454L}
observations. From {\it BeppoSAX} observations,  
\cite{2001ApJ...551..186Z} found the continuum to be well described
by a power law with $\Gamma \sim$ 1.85 $\pm$ 0.05, which is in 
close agreement with what is found from {\it NuSTAR} data
by \cite{retco}. 
The difference in the  photon index values obtained
from different sets of observations acquired from
different telescopes could point to spectral variations in the source.

Using the {\it pexrav} model
with the inclusion of a Gaussian component to account for the presence
of the Fe K$\alpha$ line in the spectrum, \cite{retco} found values of 
$\Gamma$= 1.87 $\pm$ 0.02 and E$_{cut}$ = 83$_{-8}^{+10}$ keV.
 3C 120 has been observed before by {\it BeppoSAX} and OSSE.
 By modelling the {\it BeppoSAX} data  with an e-folded power law or a thermal
Comptonization model, \cite{2001ApJ...551..186Z} found a value of  E$_{cut}$ = 150$_{-30}^{+230}$ keV. 
Using {\it ASCA} observation that was contemporaneous with an OSSE observation, 
and modelling the spectra with a broken power law multiplied
by an exponential factor, \cite{1998MNRAS.299..449W} found E$_{cut}$ = 110$_{-50}^{+130}$
keV. Within error bars, the value of E$_{cut}$ obtained from 
{\it NuSTAR} data using simple model fits matches with that known from {\it BeppoSAX} and OSSE data, however,
has improved precision, with  a manifold  reduction in 
the errors. As these observations were taken at different epochs, it 
is also likely the E$_{cut}$ is variable, but, this cannot be ascertained 
because of the large error bars in its values from earlier observations.
The Fe K$\alpha$ line 
is well fit by a Gaussian incorporated in {\it CompTT} with 
$\sigma$ of 0.29$_{-0.10}^{+0.09}$ keV  and 0.29$_{-0.10}^{+0.09}$ keV respectively 
for the slab and 
spherical geometry. This is much narrower than the value of $\sigma$
= 0.8 keV obtained from ASCA observations \citep{1997ApJ...487..636G}.

In this work, we applied physical models to the data in 
contrast to the simple phenomenological models
used earlier to understand the spectral characteristics of 3C 120 \citep{retco, 2001ApJ...551..186Z}. 
We fitted {\it CompTT}, to the observed 
spectrum and
used it to characterise the temperature and optical depth of the electrons
in the corona for two geometries, namely a sphere
and a slab.  The goodness of the fit (with a nearly identical chi-square per degree of 
freedom of 
$\chi^{2}/\nu \approx 3734/3788$) is found to be insensitive to the assumption of the coronal geometry as
 assumption of both the slab and spherical geometry fit the data 
equally well and we obtained $kT_e = 9_{-3}^{+2}$ keV for slab geometry and $kT_e = 16_{-7}^{+6}$ keV for the spherical geometry. 
 These two model assumptions about the geometry of the
corona gave different values of the optical depth  
with $\tau = 2.4_{-1.1}^{+0.6} $ and $\tau = 5.1_{-0.4}^{+0.6}$ for the slab 
and 
spherical geometry.  This
is expected  because the optical depth for a slab geometry is measured 
vertically 
while for a sphere it is measured radially.
Using {\it CompPS} an advanced Comptonization model available 
in XSPEC, we found $kT_e$ values of {25$\pm$ 2 and 26$^{+2}_{-0}$ for the
slab and spherical geometry. Within errors, these values of $kT_e$ 
matches with that obtained from the fit of the {\it EQPAIR} model
to the {\it NuSTAR} data that returned a value of $kT_e$ = 23$^{+1}_{-7}$ keV.
This value of $kT_e$ is much lower than the value 
of $kT_e$ of 176 keV  obtained by
\cite{2016MNRAS.458.2454L}. This discrepancy might be attributed to the presence of a significant
jet contribution during the epoch of observations done from INTEGRAL. 
Considering {\it CompPS} model, the derived value of $kT_e$ is 
nearly identical for both the slab and sphere geometry of the corona.
This could mean that the shape of the X-ray spectra emerging out of these
two geometries is quite similar and the available spectral data from 
{\it NuSTAR} 
is not sufficient to distinguish between these two geometries.


\subsection{Nature of the corona in 3C 120}
3C 120 is classified as a Seyfert 1 galaxy \citep{1967ApJ...149L..51B} 
and is also identified as a BLRG by \cite{1987ApJ...316..546W}. It has an 
one sided jet and is also 
known to be a $\gamma$-ray emitter in  {\it Fermi} 
data \citep{2015A&A...574A..88S,2015ApJ...799L..18T}, which provides
additional evidence for the presence of a powerful relativistic jet, already seen
in radio observations \citep{2004ApJ...615..161H}. It is known 
that BLRGs have harder X-ray spectra in comparison
to radio-quiet Seyfert galaxies \citep{2001ApJ...551..186Z}. However, spectral fits to the {\it NuSTAR}
data analysed here by \cite{retco} using {\it pexrav} model gave a photon index $\Gamma$ of 1.87 $\pm$ 0.02. This 
value is 
similar to that known for non-jetted Seyfert 1 galaxies and different from
the X-ray spectrum of AGN with relativistic jets (blazars) that have
$\Gamma$ $<$ 1.5 
\citep{2006ApJ...646...23S,2011MNRAS.411.2137G}. Though the 
derived X-ray spectral index points to negligible contribution of the jet 
emission we checked
for the signature of jets in our data by looking at the multi-wavelength
properties during the epoch of {\it NuSTAR} observations. Using the
light curves taken in the optical from the Catalina Realtime Transient
Survey (CRTS; \citealt{2009ApJ...696..870D}) and in the 15 GHz band in the radio from
the Owens Valley Radio Observatory (OVRO, 
\citealt{2011ApJS..194...29R}),
we found that 3C 120 was in a moderately low flux state during the time of {\it NuSTAR}
observation analysed here. The optical and radio 
light curves are given in Figure \ref{optical_radio} with the 
epoch of {\it NuSTAR} observations indicated as a blue dashed line.  Also, 
during the epoch of the {\it NuSTAR} observations used here, the source was
not detected in $\gamma$-rays by {\it Fermi} \citep{2015ApJ...799L..18T}. 
The $F_{var}$ for 3C 120 in the soft and hard bands are 0.065 $\pm$ 0.002 and 0.052 $\pm$ 0.003 
respectively. This is much lower than the average F$_{var}$ in X-rays shown by
the blazar class of AGN \citep{2014A&A...563A..57S,2017MNRAS.466.3309R}
Also, the variations seen in the {\it NuSTAR} data
is similar to that of Seyfert galaxies and not blazars 
\citep{2017MNRAS.466.3309R}. Model fits to the observed spectrum
using {\it CompPS} that considers thermal Comptonization gave a
value of $kT_e$ = 26$^{+2}_{-0}$ keV for a spherical geometry. On the other hand, fits
to the observed spectrum using {\it EQPAIR} that treats Comptonization from
hybrid plasma gave $kT_e$ = 23$_{-7}^{+1}$ keV. 
 Comparing CompTT and CompPS models for a spherical geometry using F-test we
find a F-value of 1.013. The test does not rule out the null hypothesis that
the two chi-square distributions are the same at the 90\% confidence level.
Between CompPS and EQPAIR model fits for a spherical geometry we find a
F-value of 1.0637, larger than the F$_{critical}$ value for a 90\% confidence,
rejecting the null hypothesis that the two chi-square distributions are the
same. As the chi-square value of CompPS matches close to unity compared to
EQPAIR, we consider CompPS model better represents the spectrum of 3C 120. 
Therefore, based on both spectral
 (presence of X-ray high energy cut-off and the X-ray photon index
being close to that known for Seyfert galaxies) and timing analysis
 (non-detection of the source in $\gamma$-rays during the epoch of
{\it NuSTAR} observations), it is clear that  the X-rays observed by
{\it NuSTAR} from 3C 120 
 is similar to that found in non-jetted Seyfert 1 galaxies
considering a model of a thermal Comptonizing corona producing the 
X-ray in 3C 120. We note that the strength of the reflection component
 obtained here showed significant differences between various model fits, 
which might the due to the low S/N of the data beyond 30 keV.

\subsection{Comparison with the  coronal properties of other AGN}
Because of the degeneracies involved in the evaluation of the 
properties of the corona from the observed X-ray spectrum, it is needed 
to simultaneously measure the 
power law slope and the cut off energy. Measurements of this demands
high quality X-ray spectra. Measurements of $E_{cut}$ 
were known for several AGN from observatories such as
{\it BeppoSAX} and INTEGRAL. However, most of these measurements
have large error bars. Recently, observations from {\it NuSTAR}
have started to provide reliable estimates of $E_{cut}$ in 
few AGN, even though it might not be sensitive to sources with $E_{cut}$ much larger
that its spectral coverage.

To compare the coronal measurements reported here for 3C 120 with that of
other AGN, we searched the literature for the availability of coronal
properties of AGN based on observations either from {\it NuSTAR} 
alone or {\it NuSTAR} observations
coupled with other telescopes. Focussing only on those sources that have $E_{cut}$ 
 measurements (with out lower limits) we arrived at a sample 
of twelve sources including 3C 120. They are given in Table \ref{log2}. 
Also, the sources listed in Table \ref{log2}
belong to different types of AGN that includes both radio-quiet Seyferts and BLRGs (3C 390.3 and 3C 120, 3C 282 and 3C 390.3). 
Analysis of a larger sample of AGN do indicate differences between
BLRGs and radio-quiet Seyfert 1 galaxies, with BLRGs having, on average
lesser Compton reflection, weaker Fe K$\alpha$ line and
harder hard X-ray spectra compared to radio-quiet Seyfert 1 
galaxies \citep{1998MNRAS.299..449W}. These differences between
BLRGs and radio-quiet Seyfert 1 galaxies are further
confirmed by \cite{2001ApJ...551..186Z},however, the authors
state that the distribution of these parameters in 
these two populations of sources is not distinct. \cite{2001ApJ...551..186Z} 
obtained mean values of $\Gamma$ = 1.74 $\pm$ 0.04 and 1.95 $\pm$ 0.05 
for BLRGs and radio-quiet Seyferts respectively. The value of $\Gamma$
obtained for 3C 120 by \cite{retco}
 is closer to what is known for radio-quiet Seyfert 1 galaxies and is steeper
than the other two BLRGs 3C 282 and 3C 390.3. This also supports the 
dormant state of the jet of 3C 120 during the epoch of {\it NuSTAR} observations
reported here. Though the $kT_e$ values of 3C 120 and 3C390.3 agree within a factor of two,
the value of $kT_e$ obtained for 3C 382 another BLRG is much larger.
Therefore, based on existing data from {\it NuSTAR}, it
is very difficult to say if the coronal properties of 
radio-loud AGN (BLRGs) and radio-quiet AGN (radio-quiet Seyfert 1 galaxies)
are similar or different. Understanding the connection between
radio-emission and coronal properties if any needs observations
on a large number of sources of both types analysed in a
homogeneous manner. For this modest sample of sources with {\it NuSTAR} observations
culled from literature, we looked for correlation
of $kT_e$ with other physical parameters of the sources such as
$\Gamma$ and the black hole mass. No correlation
could be established (Figure \ref{corr}). Therefore, more and more measurement of $kT_e$
on a large
sample that comprises both radio-loud and radio-quiet AGN  
are needed to know for the existence or absence of such 
correlations and largely to better 
understand the nature of the corona in AGN.

\begin{figure}
\centering
 \includegraphics[width=8.5cm,height=7.5cm]{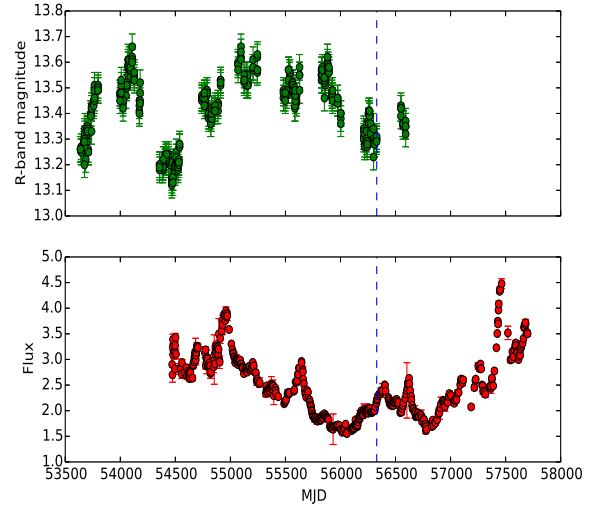}
\caption{\label{optical_radio}Long term optical V-band light curves from CRTS 
(top panel) and 15 GHz radio light curves from OVRO (bottom panel). The
epoch of {\it NuSTAR} observation studied here is indicated by the 
blue dashed line.}
\end{figure}

\begin{figure}
\centering
\vbox{
   \includegraphics[width=8.5cm,height=7.5cm]{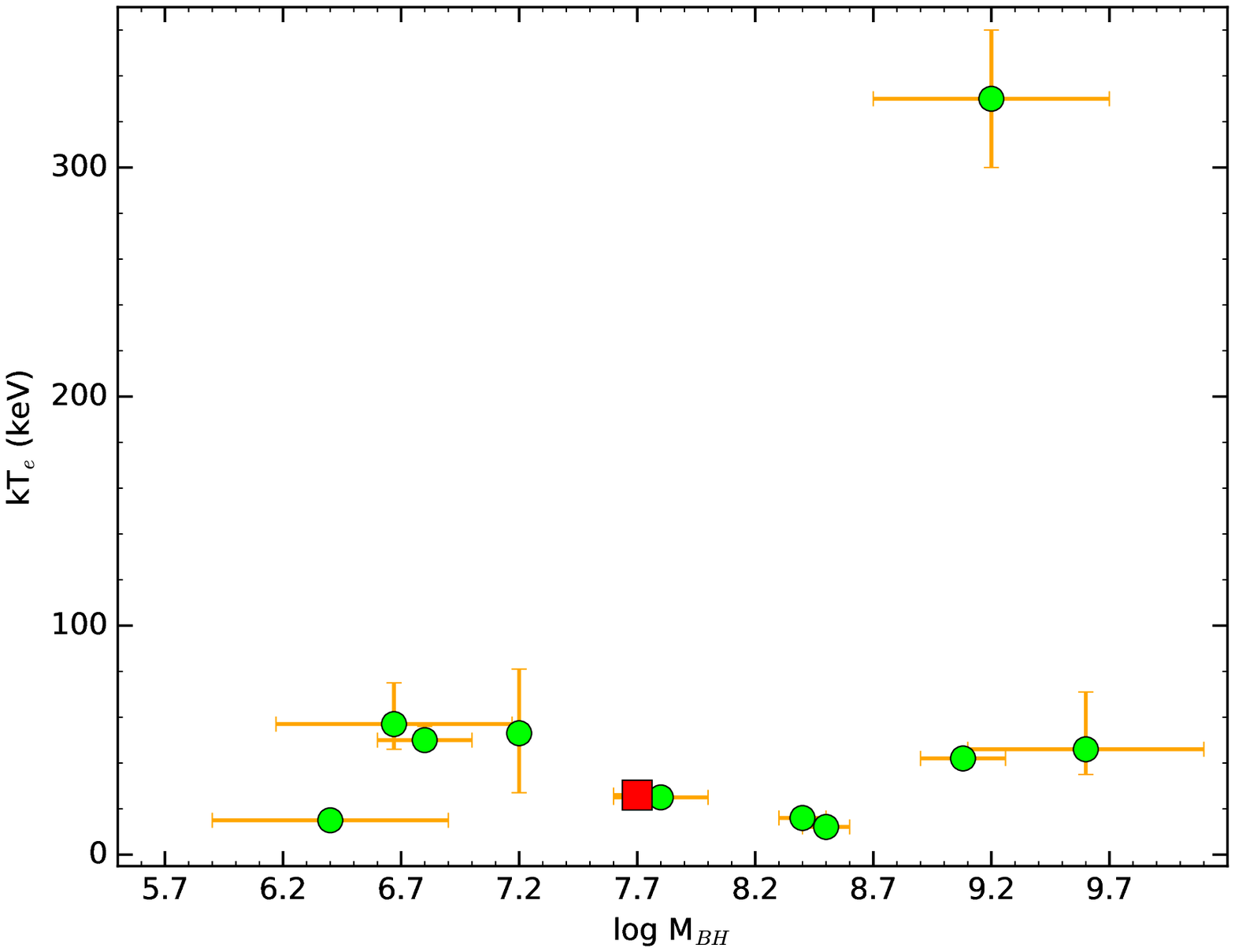}
   \includegraphics[width=8.5cm,height=7.5cm]{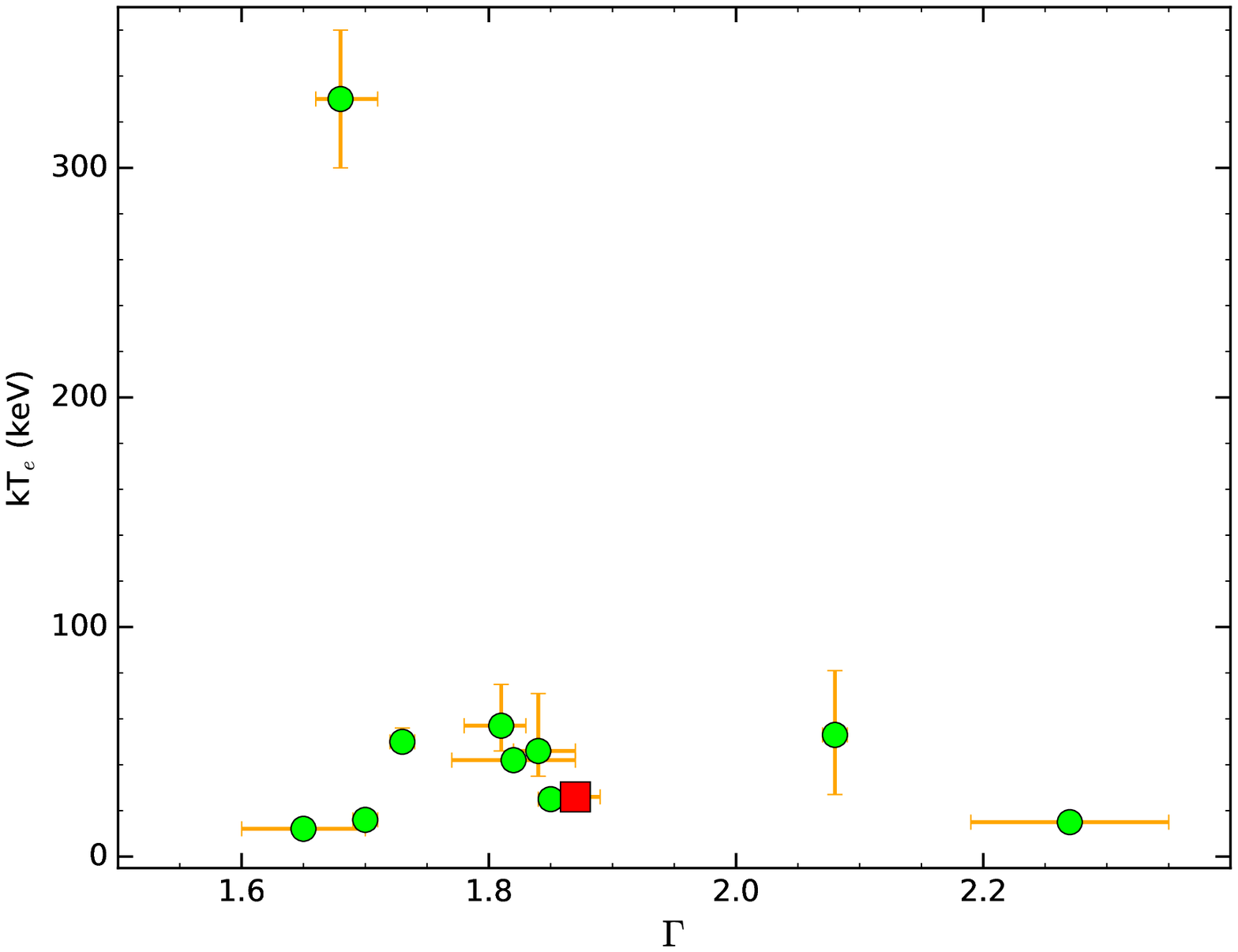}
    }
\caption{\label{corr}Correlation between $kT_e$ in keV  
and logarithm of M$_{BH}$ in units of M$_{\odot}$ (top panel) 
and $\Gamma$ (bottom panel) for the sample of sources culled from 
literature with known $E_{cut}$ measurements from {\it NuSTAR} (without lower
limits). 3C 120 studied in this work is shown as a red square}
\end{figure}

\section{Summary}
We have carried out timing and spectral analysis of the Seyfert 1 galaxy
3C 120 using $\sim$120 ks observations from {\it NuSTAR}. The results of our 
analysis are summarized below:
\begin{enumerate}
\item The source showed variations in the count rates during the duration of the X-ray 
observations. The amplitude of count rate variations characterized by F$_{var}$ are
found to be 0.065 $\pm$ 0.002 and 0.052 $\pm$ 0.003 for the soft 
(3 $-$ 10) keV  and the hard ($10-79$) keV  bands respectively, thus 
showing more variations in the soft band relative to the hard band

\item The X-ray spectrum characterised by HR is found not to show a  
correlation with the flux variations in the total band, indicating 
the spectrum was non-variable during the epoch of {\it NuSTAR}
observation.

\item From fit of {\it CompTT} model to the time averaged spectrum we found
evidence for the presence of weak Fe K$\alpha$ line in the data at 6.4 keV with an 
equivalent width of 60 $\pm$ 5 eV.
The line is best fit by a Gaussian with a $\sigma$ of 0.29 keV


\item Using the Comptonization model {\it CompPS} to fit the observed 
spectrum, we derived the kinetic temperature of the coronal electrons to be 
$kT_e$ = 25 $\pm$ 2 keV  with a Compton y parameter of 
$ y$ = 2.2 $\pm$ 0.2 for a slab geometry. 
This is similar to the
value of the kinetic temperature of $kT_e$ = 26$^{+2}_{-0}$ keV
obtained for a spherical 
geometry with a $y$ of 2.99$^{+2.99}_{-0.18}$. Also, fitting the observed
spectrum with {\it EQPAIR} gave
a best fit value of $kT_e$ = 23$^{+1}_{-7}$ keV.
Thus, fits to the data 
with the two most advanced 
Comptonization models available in {\it XSPEC} namely {\it CompPS} and
{\it EQPAIR}  gave similar values of coronal 
temperature. 
 It is likely that the electrons participating in the comptonization
process is predominantly thermal. Comptonization by non-thermal electrons
if any is in-significant as (i) the source is not detected in $\gamma$-rays
during the epoch of {\it NuSTAR} observations and (ii) the X-ray photon
index is similar to that known for Seyfert galaxies

\item 3C 120 is known to have a large scale radio jet and is  
also a $\gamma$-ray emitter. 
However, {\it NuSTAR} data analysed here has made possible the detection of
coronal spectral signatures, constrain $kT_e$ and the reflection
features, which are found  
similar to that known for radio-quiet 
Seyfert galaxies. This indicates
that the contribution of jet emission to the X-ray is negligible in the 
{\it NuSTAR} data and is likely to be weak during the epoch of {\it NuSTAR}
observations. Additional support to this is provided by similar value of $kT_e$ obtained
by both {\it CompPS} and {\it EQPAIR} model fits to {\it NuSTAR} observations.
This is also supported by the low/moderate radio and optical flux states as well
as non-detection by {\it Fermi} during the epoch of {\it NuSTAR} observations.
To constrain the contribution of jet emission if any to the X-ray emission 
from 3C 120 requires observations at energies higher than that covered
by {\it NuSTAR}.

\end{enumerate}
 
 




%

\acknowledgments
We thank the referee for his/her critical and valuable comments that
improved the paper significantly.
This research made use of data from the {\it NuSTAR} mission, a project 
led by the 
California Institute of Technology, managed by the Jet Propulsion Laboratory, 
and funded by NASA, XMM-Newton, an ESA science mission with instruments and 
contributions directly funded by ESA Member States and NASA. This research has 
made use of the {\it NuSTAR} Data Analysis Software (NuSTARDAS) jointly developed by 
the ASI Science Data Center (ASDC, Italy) and the California Institute of 
Technology (USA).

\bibliographystyle{apj}
\bibliography{master.bib}

\end{document}